\documentclass[a4paper,11pt]{article}
\usepackage{pos}

\title{Axion superradiance}

\author[a]{Francesca Chadha-Day}

\affiliation[a]{Institute for Particle Physics Phenomenology\\
Department of Physics\\
Durham University\\
Durham\\
United Kingdom}

\emailAdd{francesca.chadha-day@durham.ac.uk}

\abstract{Light bosonic fields may suffer an instability around a rotating compact object. This process, known as superradiance, leads to the exponential amplification of the field around a black hole or neutron star, while the spin of the central object is correspondingly depleted. The discovery of a highly spinning black hole could therefore be used to constrain the existence of light bosons such as axions in a particular range of masses. These constraints apply for very low non-gravitational couplings between the boson and the Standard Model, offering a powerful search strategy for new physics. However, care must be taken to include the more complex effects of the black hole's astrophysical environment. Conversely, stellar superradiance could allow us to probe additional non-gravitational interactions between a new boson at the stellar matter. In this article, I will discuss the current status and future directions of axion superradiance. This is a contribution to the proceedings of the 3rd General Meeting of the COST Action COSMIC WSIPers.\\
\hfill \\
IPPP/25/94}

\FullConference{

3rd General Meeting of the COST Action COSMIC WSIPers (CA21106) (COSMICWISPers2025)
9–12 Sept 2025\\
Sofia, Bulgaria and Annecy, France\\
}

\begin{document}
\maketitle

\section{Introduction}

Radiation incident on a dissipative system may be either depleted or amplified. The latter case is known as superradiance. Superradiance was first detected in the laboratory with water surface waves incident on a draining vortex \cite{1612.06180}. The well known phenomenon of Cherenkov radiation, in which radiation is emitted from a charged particle whose velocity $v$ is greater than the radiation's phase velocity $v_{\rm ph}$ in medium, can also be understood as a form of superradiance \cite{gr-qc/9803033}. An analgous effect may occur when radiation is incident on a rotating body even in vacuum. This is because the {\it angular} phase velocity of radiation with azimuthal quantum number $m$ and angular frequencey $\omega$ is given by $v_{\phi} = \omega/m$, which may be lower than the angular velocity $\Omega$ of the rotating body. Indeed, in the cases of black hole and stellar superradiance discussed below, we will see that superradiance occurs when $\Omega > v_{\phi} \implies\omega < m \Omega$.\\

In this article, we will give a brief, pedagogical discussion the state of the art in black hole and stellar superradiance and their potential to detect axions and other physics Beyond the Standard Model.

\section{Black hole superradiance}

A black hole (BH) spacetime is a dissipative system due to the presence of the horizon. It is well known that a rotating (Kerr) BH will amplify radiation that passes through its ergoregion. This can be understood as a superradiant scattering process. Furthermore, massive radiation may be trapped in the vicinity of the BH, in effect passing through the ergoregion numerous times. This trapped radiation may therefore be exponentially amplified in a {\it superradiant instability}. 

To calculate the growth rate of this superradiant instability, we must consider bound states of some bosonic field $\phi$ around a Kerr BH. (The Pauli exclusion principle prevents a superradiant instability for fermionic fields.) Except for very close to the BH, it is sufficient to use a Newtonian gravitational potential, and the system therefore resembles a hydrogen atom with bound states $\psi_{nlm}(r)$. However, we find that in the Kerr BH spacetime the corresponding eigen-energies $\omega_{nlm}$ have an imaginary component, corresponding either to $\phi$ falling into the BH or to the superradiant growth of $\phi$.

We will first consider the evolution of a scalar field $\phi$ of mass $\mu$ around a rotating BH. This is described by the action

\begin{equation}
S = \int d^4 x \sqrt{-g} (- \frac{1}{2} \triangledown_{\mu} \phi \triangledown^{\mu} \phi - \frac{1}{2} \mu \phi^2),
\end{equation}

where the metric determinant $g$ and covariant derivatives are derived from the Kerr metric. We impose ingoing boundary conditions at the horizon. The resulting equations of motion admit quasi-bound states with $\omega_{nlm} = \omega_R+i\omega_I$, where $\omega_R \lesssim \mu$ and $\omega_I$ can be found numerically and in some cases analytically. $\omega_I > 0$ corresponds to superradiance with an e-folding timescale $\tau \sim 1/\omega_I$. Note that this approach neglects the backreaction of the field's evolution onto the BH. However,  it is clear from conservation of energy and angular momentum that the BH spins down as the field is amplified. 

The solution for $\omega_I$ and therefore the superradiance rate behaves as follows \cite{0705.2880}:

\begin{itemize}
\item A superradiant instability exists when $\omega_R < m \Omega$, where $m$ is the azimuthal quantum number of the bound state and $\Omega$ is the angular velocity of the BH.
\item The instability is most efficient when $G M \mu \sim 1$, where $M$ is the mass of the BH i.e. when the BH's gravitational radius is similar to the field's Compton radius.
\item The instability is less efficient for higher $l$ and $m$ modes.
\end{itemize}

These results tell us that efficient superradiance is not possible for Standard Model bosons. The electroweak bosons are much too heavy for observable superradiance with either astrophysical or supermassive BHs. Detailed study of BH superradiance involving a photon with a plasma mass suggests that this is also probably not observable \cite{2306.12490}. However, superradiance may be a highly effective probe of light Beyond the Standard model bosons such as axions and dark photons, as first noted in \cite{Arvanitaki:2009fg}.

We have so far demonstrated that the amplitude of a light bosonic field $\phi$ around a rotating BH may grow exponentially. To observe this effect, we require some initial non-zero value for $\phi$ close to the BH. The standard lore is that this can be provided by a quantum fluctuation, rendering the superradiant instability inevitable for any bosonic degree of freedom with appropriate mass. This possibility relies on some assumptions about quantum gravity. The superradiant instability could also be seeded by a cosmological abundance of $\phi$, for example as (a fraction of) dark matter or dark energy. However, we must note that if the initial seed is $\mathcal{O}(1 \%)$ of the BH's mass with both superradiant and nonsuperradiant modes populated (corresponding to bound states with both positive and negative $m$), the superradiant instability may be suppressed \cite{1812.02758}. \\

We now turn to the question of how we might observe the superradiant cloud of $\phi$ quanta given that they presumably interact only very weakly with the Standard Model. The first possibility is the observation of a `bosenova' explosion when the cloud collapses due to the self-interactions of $\phi$ \cite{1004.3558}. Depending on the other interactions of $\phi$, it may also be possible to detect the cloud indirectly, for example via mono-energetic gravitational wave emission from transitions between bound states \cite{1604.03958}, birefringence \cite{1711.08298}, electromagnetic emission via lasing \cite{1811.04950}, or the effect on orbtis in binary systems \cite{1910.06977}.

The presence of the superradiant cloud may be detected in a more model independent way by searching for BH spin depletion. BH spins can be inferred from analysis of the X-ray spectra of BH binaries and of gravitational wave emission from BH mergers. The discovery of a highly spinning BH of mass $M$ suggests that no efficient superradiant instability exists for this mass. Conversely, superradiance would lead to gaps in the BH mass vs spin plot. Note that bosons whose self-interactions are too strong cannot be probed in this way, as the cloud would become unstable to bosenova explosion before a significant fraction of the BH's spin had been depleted. X-ray and gravitational wave spin measurements have recently been used to place constraints on ultra-light bosons in \cite{2406.10337} and \cite{2507.21788} respectively. These searches are sensitive to bosons with masses in the range $10^{-19} \, {\rm eV} \lesssim \mu \lesssim 10^{-18} \, {\rm eV}$ with decay constant $f \gtrsim 10^{14} \, {\rm GeV}$ and bosons with masses in the range $10^{-13} \, {\rm eV} \lesssim \mu \lesssim 10^{-12} \, {\rm eV}$ with decay constant $f \gtrsim 10^{12} \, {\rm GeV}$. The former case corresponds to supermassive BHs and the latter to astrophysical BHs. Note that the bosons' self interactions are of strength $\sim f^{-2}$. These searches provide a useful complement to other searches for new bosons that are typically sensitive to low decay constants.

So far, we have considered a scalar field evolving in the background of an isolated BH. Real BHs have accretion disks. The effect of the accretion disk on superradiance has recently been considered in \cite{2404.09955} for the case of a supermassive BH at the core of an active galactic nucleus. Even considering only gravitational interactions of the superradiant scalar, the presence of the accretion disk significantly increases the complexity of the superradiant system. In this case, the evolution of the mass and angular momentum of the BH has contributions from both superradiance and accretion. Meanwhile, the evolution of the mass and angular momentum of the superradiant cloud depends on both the superradiance time scale itself and on gravitational wave emission of the cloud, which may be significantly enhanced due to the dynamics of the accretion disk according to the results of \cite{2404.09955}. The rate of accretion itself also depends on the mass and spin of the BH. A system of coupled ODEs describing the evolution of the BH's mass and spin and of the disk luminosity are therefore obtained. This raises the exciting possibility that readily available measurements of AGN luminosoties could be used to search for superradiance. We must also note that the existience of mechanisms such as accretion and mergers to spin up black holes must be taken into account when placing bounds on light bosons from the observation of single highly spinning BHs.

\section{Stellar superradiance}

It is well known that rotating stars can also exhibit superradiant scattering and host superradiant clouds \cite{Zeldovich}. As stars do not possess a horizon, stellar superradiance relies on additional non-gravitational interactions between the boson and the stellar material to provide the required dissipative dynamics. This dissipative interaction becomes amplifying in the superradiant regime due to the star's rotation.

The majority of studies of stellar superradiance focus on neutron stars, as superradiance is less efficient for less compact objects. For example, neutron star superradiance has been studied in the context of dark photons interacting with the star via a hidden sector conductivity \cite{1704.06151}. Stellar superradiance can also occur in the context of more complex systems. For example, \cite{1904.08341} demonstrated that superradiance can occur due to axion-photon mixing in the magnetosphere of a neutron star. The photon's Standard Model interaction with the neutron star magnetosphere provides the required dissipation, while the axion component remains trapped in a bound state around the neutron star due to its mass. Due to the mixing between the axion and photon in the neutron star's strong magnetic field, the axion's bound state energy has an imaginary component exhibiting superradiant amplification. The rate of this process depends on modeling of the photon's conductivity in the magnetosphere, an issue that is the subject of ongoing debate \cite{2503.19978,2504.04209}. In any case, this system demonstrates the diversity of possibilities we must consider to study stellar superradiance.

Many different Beyond the Standard Model (BSM) interactions could lead to stellar superradiance. Furthermore, stellar environments are complex, with many more degrees of freedom than black holes. Stellar superradiance therefore offers the potential to observe a wide range of new physics, in particular as the spin-down of neutron stars can be observed directly. This motivates the development of a more general method for computing stellar superradiance rates from a given BSM Lagrangian, allowing a more systematic search than previous work that has focused on solving particular systems. Such a method was presented in \cite{2207.07662}, combining insights from thermal field theory and worldline effective field theory (worldline EFT). \\

As discussed in \cite{2207.07662}, stellar superradiance depends on the {\it net} damping rate $\Gamma_\phi$ of the field $\phi$ into the star (i.e. the difference between the absorption and emission rates). This appears in the classical equation of motion as

\begin{equation}
\label{eq:classical}
\partial^2 \phi+\mu^2 \phi+\Gamma_\phi \dot{\phi}=0.
\end{equation}

Using thermal field theory, we can show that this is given by:

\begin{equation}
\Gamma_\phi = -\lim _{p \rightarrow 0} \operatorname{Im} \Pi(p) / p^0,
\end{equation}

where $\Pi$ is the self energy of $\phi$ in the thermal bath of the star, characterized by a temperature $T$ and chemical potentials $\mu_i$ for all relevant degrees of freedom, such as fundamental particles or plasma modes. This is evaluated in the long wavelength limit appropriate for the non-relativistic particles in bound states around the star. As discussed in \cite{2512.13816}, great care must be taken over this point. The long wavelength damping rate cannot be naively extrapolated from the short wavelength damping rate appropriate for stellar cooling calculations. In particular, \cite{2512.13816} identifies two effects in the long wavelength regime that suppress stellar superradiance - collective scattering effects and non-relativstic gradient couplings. 

To relate this damping rate to the superradiant instability, we employ the worldline effective field theory description of superradiance developed in \cite{1609.06723}. We describe the interaction of $\phi$ and the star by expanding in $\frac{R}{\lambda}$, where $R$ is the star's radius and $\lambda$ is the Compton wavelength of $\phi$. The extended nature of the star is thus described by an infinite series of interactions between $\phi$ and a point-like object, given by the interaction Hamiltonian:

\begin{equation}
\begin{aligned}
H_{\mathrm{int}}(t, \mathbf{x})=& \partial^I \phi(x) \mathcal{O}_I^{(1)}(x) \delta^{(3)}(\mathbf{x}-\mathbf{y}(t))+\\
& \partial^I \partial^J \phi(x) \mathcal{O}_{I J}^{(2)}(x) \delta^{(3)}(\mathbf{x}-\mathbf{y}(t))+\ldots,
\end{aligned}
\end{equation}

where $\mathbf{y}(t)$ is the worldline of the star and $\mathcal{O}$ are the worldline EFT operators. The advantage of this approach is that the rotation of the star can be modelled by simply rotating these operators with 3D rotation matrices $R_I^J$.

\begin{equation}
\begin{aligned}
H_{\mathrm{int}}(t) &=\partial^I \phi(t) R_I^J(t) \mathcal{O}_J^{(1)}(t) \\
&+\partial^I \partial^J \phi(t) R_I^K(t) R_J^L(t) \mathcal{O}_{K L}^{(2)}(t)+\ldots,
\end{aligned}
\end{equation}

where we have also specialised to the rest frame of the star. We can now calculate the superradiance scattering rate from $H_{\mathrm{int}}(t)$. We first calculate the absorption probability for the state $|\omega, \ell, m \rangle $:

\begin{equation}
P_{\mathrm{abs}}=\sum_{X_f} \frac{\left|\left\langle X_f ; 0|S| X_i ; \omega, \ell, m\right\rangle\right|^2}{\langle\omega, \ell, m \mid \omega, \ell, m\rangle},
\end{equation}

where $X_i$ and $X_f$ are the states of the star before and after the absorption process and $S$ is the action derived from $H_{\mathrm{int}}(t)$. We can similarly compute the emission probability and hence the superradiant scattering rate $Z_{\ell m}$. We find:

\begin{equation}
\label{eq:scattering}
Z_{\ell m} = \frac{\Phi_{\mathrm{out}}-\Phi_{\mathrm{in}}}{\Phi_{\mathrm{in}}} =\frac{\ell ! q^{2 \ell+2}}{4 \pi(2 \ell+1) ! ! v \omega} \rho_{\ell}(m \Omega-\omega),
\end{equation}

where $\Phi_{\mathrm{in}}$ and $\Phi_{\mathrm{out}}$ are the ingoing and outgoing fluxes, $q$ is the amplitude of the wave's spatial momentum and $v$ is its group velocity. Crucially, $\rho_{\ell}(m \Omega-\omega)$ is related to the worldline field theory operators and hence must be found with a matching calculation, as per the standard approach to EFTs.

To compute the rate of the superradiant instability, we must now consider bound states. For example, the absorption probability for the bound state $ \mid nlm \rangle$ around the star is computed as

\begin{equation}
P_{\mathrm{abs}}=\sum_{X_f}\left|\left\langle X_f ; 0|S| X_i ; n \ell m\right\rangle\right|^2.
\end{equation}

We can similarly compute the emission probability for the bound state and hence the growth rate of the superradiant instability

\begin{equation}
\label{eq:instability}
\Gamma_{n \ell m} = \Gamma_{\mathrm{em}} - \Gamma_{\mathrm{abs}} =\frac{A_{n \ell m}}{2 \omega_{\ell n}}\left(\frac{1}{r_{n \ell}}\right)^{2 \ell+3} \rho_{\ell}\left(m \Omega-\omega_{\ell n}\right),
\end{equation}

where $r_{n\ell} = (n+\ell+1)/(2GM \mu^2)$ and $A_{n \ell m}$ is a combinatoric factor given in \cite{2207.07662}.

To obtain the full expression for $\Gamma_{n \ell m}$ we must obtain $\rho_{\ell}(m \Omega-\omega_{\ell n})$ via a matching calculation. We match the superradiant scattering rate calculated from the worldline EFT given in equation \eqref{eq:scattering} to the same quantity calculated from the classical equations of motion \cite{1505.05509}. By matching these results we can obtain $\rho_{\ell}(m \Omega-\omega_{\ell n})$ (or equivalently the EFT coefficients). We can then substitute these into equation \eqref{eq:instability} to obtain the rate of the superradiant instability.

This gives a superradiance rate:

\begin{align}
\label{eq:SRrate}
    &\Gamma_{n \ell m} = C_{n \ell m} \left(\frac{R}{r_{n\ell}} \right)^{(2\ell +3)}  \frac{(m \Omega- \omega_{\ell n})}{\omega_{\ell n}} \Gamma_\phi,
\end{align}

where we recall that the damping $\Gamma_{\phi}$ of the field $\phi$ into the star can be calculated in thermal field theory and $C_{n \ell m}$ is a combinatoric factor given in \cite{2207.07662}. This allows us to calculate the stellar superradiance rate for any interaction between a bosonic field and a star. Indeed, equation \eqref{eq:SRrate} can be used to calculate the superradiance rate for any bosonic field $\phi$ once the damping rate $\Gamma_\phi$ of $\phi$ into the star has been computed. It is not necessary to perform further calculations using matching or the worldline effective field theory for each boson, but this approach allows us to understand both black hole and stellar superradiance in terms of the rotation of operators describing the interactions between the boson and the BH or star.

\section{Summary and outlook}

Black hole superradiance offers a purely gravitational probe of Beyond the Standard Model bosons such as axions. Stellar superradiance can probe additional interactions between new bosons and the Standard Model, and in particular is sensitive to the long wavelength limit of these interactions. This poses challenges in calculating the correct damping rate given our ignorance of the macroscopic degrees of freedom of neutron star matter. While the theoretical underpinning of both black hole and stellar superradiance is now well understood, further work is needed to incorporate the complications of astrophysics into both of these processes.

\section*{Acknowledgments}
\noindent This article is based upon a talk and discussions at the 3rd General Meeting of the COST Action COSMIC WSIPers (CA21106), supported by COST (European Cooperation in Science and Technology. FCD is supported by STFC consolidated grant ST/X003167/1.


\begin{thebibliography}{99}



\bibitem{1612.06180}
T.~Torres, S.~Patrick, A.~Coutant, M.~Richartz, E.~W.~Tedford and S.~Weinfurtner,
Nature Phys. \textbf{13} (2017), 833-836
doi:10.1038/nphys4151
[arXiv:1612.06180 [gr-qc]].

\bibitem{gr-qc/9803033}
J.~D.~Bekenstein and M.~Schiffer,
Phys. Rev. D \textbf{58} (1998), 064014
doi:10.1103/PhysRevD.58.064014
[arXiv:gr-qc/9803033 [gr-qc]].

\bibitem{0705.2880}
S.~R.~Dolan,
Phys. Rev. D \textbf{76} (2007), 084001
doi:10.1103/PhysRevD.76.084001
[arXiv:0705.2880 [gr-qc]].

\bibitem{2306.12490}
E.~Cannizzaro, F.~Corelli and P.~Pani,
Phys. Rev. D \textbf{109} (2024) no.2, 023007
doi:10.1103/PhysRevD.109.023007
[arXiv:2306.12490 [gr-qc]].

\bibitem{Arvanitaki:2009fg}
A.~Arvanitaki, S.~Dimopoulos, S.~Dubovsky, N.~Kaloper and J.~March-Russell,
Phys. Rev. D \textbf{81} (2010), 123530
doi:10.1103/PhysRevD.81.123530
[arXiv:0905.4720 [hep-th]].

\bibitem{1812.02758}
G.~Ficarra, P.~Pani and H.~Witek,
Phys. Rev. D \textbf{99} (2019) no.10, 104019
doi:10.1103/PhysRevD.99.104019
[arXiv:1812.02758 [gr-qc]].


\bibitem{1004.3558}
A.~Arvanitaki and S.~Dubovsky,
Phys. Rev. D \textbf{83} (2011), 044026
doi:10.1103/PhysRevD.83.044026
[arXiv:1004.3558 [hep-th]].

\bibitem{1604.03958}
A.~Arvanitaki, M.~Baryakhtar, S.~Dimopoulos, S.~Dubovsky and R.~Lasenby,
Phys. Rev. D \textbf{95} (2017) no.4, 043001
doi:10.1103/PhysRevD.95.043001
[arXiv:1604.03958 [hep-ph]].

\bibitem{1711.08298}
A.~D.~Plascencia and A.~Urbano,
JCAP \textbf{04} (2018), 059
doi:10.1088/1475-7516/2018/04/059
[arXiv:1711.08298 [gr-qc]].

\bibitem{1811.04950}
T.~Ikeda, R.~Brito and V.~Cardoso,
Phys. Rev. Lett. \textbf{122} (2019) no.8, 081101
doi:10.1103/PhysRevLett.122.081101
[arXiv:1811.04950 [gr-qc]].

\bibitem{1910.06977}
M.~Kavic, S.~L.~Liebling, M.~Lippert and J.~H.~Simonetti,
JCAP \textbf{08} (2020), 005
doi:10.1088/1475-7516/2020/08/005
[arXiv:1910.06977 [astro-ph.HE]].

\bibitem{2406.10337}
S.~Hoof, D.~J.~E.~Marsh, J.~Sisk-Reyn{\'e}s, J.~H.~Matthews and C.~Reynolds,
doi:10.1093/mnras/staf1564
[arXiv:2406.10337 [hep-ph]].

\bibitem{2507.21788}
A.~Caputo, G.~Franciolini and S.~J.~Witte,
[arXiv:2507.21788 [hep-ph]].

\bibitem{2404.09955}
P.~Sarmah, H.~Verma, K.~Cheung and J.~Silk,
Mon. Not. Roy. Astron. Soc. \textbf{538} (2025) no.2, 943-962
doi:10.1093/mnras/staf326
[arXiv:2404.09955 [astro-ph.HE]].

\bibitem{Zeldovich}
Y. ~B. ~Zel'Dovich,
ZhETF Pisma Redaktsiiu \textbf{14} (1971) 270.


\bibitem{1704.06151}
V.~Cardoso, P.~Pani and T.~T.~Yu,
Phys. Rev. D \textbf{95} (2017) no.12, 124056
doi:10.1103/PhysRevD.95.124056
[arXiv:1704.06151 [gr-qc]].

\bibitem{1904.08341}
F.~V.~Day and J.~I.~McDonald,
JCAP \textbf{10} (2019), 051
doi:10.1088/1475-7516/2019/10/051
[arXiv:1904.08341 [hep-ph]].

\bibitem{2503.19978}
T.~F.~M.~Spieksma and E.~Cannizzaro,
doi:10.1088/1475-7516/2025/06/028
[arXiv:2503.19978 [hep-ph]].

\bibitem{2504.04209}
T.~Sirki{\"a}, M.~Heikinheimo and K.~Tuominen,
Phys. Rev. D \textbf{111} (2025) no.11, 115030
doi:10.1103/gs6h-xlk8
[arXiv:2504.04209 [hep-ph]].

\bibitem{2207.07662}
F.~Chadha-Day, B.~Garbrecht and J.~McDonald,
JCAP \textbf{12} (2022), 008
doi:10.1088/1475-7516/2022/12/008
[arXiv:2207.07662 [hep-ph]].

\bibitem{2512.13816}
Z.~Bai, V.~Cardoso, Y.~Chen, Y.~Li, J.~I.~McDonald and H.~Seong,
[arXiv:2512.13816 [hep-ph]].

\bibitem{1609.06723}
S.~Endlich and R.~Penco,
JHEP \textbf{05} (2017), 052
doi:10.1007/JHEP05(2017)052
[arXiv:1609.06723 [hep-th]].



\bibitem{1505.05509}
V.~Cardoso, R.~Brito and J.~L.~Rosa,
Phys. Rev. D \textbf{91} (2015) no.12, 124026
doi:10.1103/PhysRevD.91.124026
[arXiv:1505.05509 [gr-qc]].



\end{thebibliography}
\end{document}